\documentclass[a4paper,11pt]{article}
\usepackage{pos}
\usepackage{multirow}
\usepackage{subcaption}
\usepackage{watermark}

\title{Robust constraints on Lorentz Invariance Violation from H.E.S.S., MAGIC and VERITAS data combination}
 \ShortTitle{Robust constraints on LIV with H+M+V data combination}

\author*[1,2]{Christelle Levy}
\author[1]{Julien Bolmont}
\author[3]{Sami Caroff}
\author[4]{Markus Gaug}
\author[5]{Alasdair Gent}
\author[1]{Agnieszka Jacholkowska}
\author[6]{Daniel Kerszberg}
\author[7]{Tony T.Y. Lin}
\author[6]{Manel Martinez}
\author[6]{Leyre Nogués}
\author[5]{A. Nepomuk Otte}
\author[8]{Cedric Perennes}
\author[1]{Michele Ronco}
\author[9]{Tomislav Terzi\'c}

\affiliation[1]{Sorbonne Université, CNRS, IN2P3, Laboratoire de Physique Nucléaire et de Hautes Energies (LPNHE), 4 place Jussieu, F-75252 Paris, France}

\affiliation[2]{Laboratoire Univers et Théorie (LUTh), Observatoire de Paris-Meudon, 5 Place Jules Janssen, 92190 Meudon, France}

\affiliation[3]{Laboratoire d'Annecy de Physique des Particules, CNRS, 9 chemin de Bellevue, Annecy, France}

\affiliation[4]{Unitat de Física de les Radiacions, Departament de Física, and CERES-IEEC, Universitat Autònoma de Barcelona E-08193 Bellaterra, Spain}

\affiliation[5]{Georgia Tech, Atlanta, U.S.A.}

\affiliation[6]{Institut de Física d’Altes Energies (IFAE), The Barcelona Institute of Science and Technology (BIST), Barcelona, Spain}

\affiliation[7]{McGill University, Montreal, Canada}

\affiliation[8]{Università di Padova and INFN, I-35131 Padova, Italy}
 
\affiliation[9]{University  of  Rijeka,  Department  of  Physics, 51000  Rijeka,  Croatia}

\emailAdd{clevy@lpnhe.in2p3.fr}
\emailAdd{jbolmont@lpnhe.in2p3.fr}

\abstract{Gamma-Ray bursts, flaring active galactic nuclei and pulsars are distant and energetic astrophysical sources, detected up to tens of TeV with Imaging Atmospheric Cherenkov Telescopes (IACTs). Due to their high variability, they are the most suitable sources for energy-dependent time-delay searches related to Lorentz Invariance Violation (LIV) predicted by some Quantum Gravity (QG) models. However, these studies require large datasets. A working group between the three major IACTs ground experiments - H.E.S.S., MAGIC and VERITAS - has been formed to address this issue and combine for the first time all the relevant data collected by the three experiments in a joint analysis.

This proceeding will review the new standard combination method. The likelihood technique used to deal with data from different source types and instruments will be presented, as well as the way systematic uncertainties are taken into account. The method has been developed and tested using simulations based on published source observations from the three experiments. From these simulations, the performance of the method will be assessed and new light will be shed on time delays dependencies with redshift.}

\FullConference{37$^{\rm{th}}$ International Cosmic Ray Conference (ICRC 2021)\\
		July 12th -- 23rd, 2021\\
		Online -- Berlin, Germany}

\begin{document}
\maketitle

\section{Introduction}
\label{sec:intro}
Although it is notoriously difficult to extract observable predictions from quantum gravity (QG) current models, departures from Lorentz invariance predicted by some of them (see \citep{Gamibini1999,Mavromatos2010}) have become one of the rare observable features we could expect. Lorentz invariance could be modified at energies approaching the Planck scale ($E_P = \sqrt{\hbar c^5 / G}\, \simeq 10^{19}$ GeV) where General Relativity (GR) and quantum mechanics (QM) should compete, while retaining the symmetry at lower energies. Departures from Lorentz invariance through violation (noted LIV for Lorentz Invariance Violation) or deformations can be taken into account with a modified dispersion relation for photons in vacuum such as \citep{Amelino1998}: 
\begin{equation}
\label{eq:disprel1}
E^2 \simeq p^2 c^2\times\left[1 \pm \sum_{n=1}^\infty \left(\frac{E}{E_{QG}}\right)^n\right],
\end{equation}
where $c$ is the low energy limit of the speed of light, $n$ is the correction order and $E_{QG,n}$ the energy scale of QG effects expected to be of the order of the Planck scale $E_P$. The sign $\pm$ allows for so-called subluminal (+) or superluminal (-) effects. Considering the sensitivity of current detectors, only linear $n=1$ or quadratic $n=2$ modifications are of interest for experimental searches. We introduce the notation $E_{QG,n}$ to reflect the fact that LIV analyses have different sensitivites for these two correction orders.

These quantum-spacetime effects being cumulative, very distant astrophysical sources are used to compensate for the smallness of the effect ($E/E_{QG,1} \propto 10^{-19} - 10^{-14}$) \citep{Amelino1998, Liberati2006}. The overall effect could become detectable in the form of energy-dependent time delays in the light curves as emitted photons travel large distances. Variable or transient sources such as gamma-ray bursts (GRBs), flaring active galactic nuclei (AGNs) and pulsars (PSRs) form a group of suitable candidates for LIV studies.

From Equation~(\ref{eq:disprel1}), it can be shown the group velocity becomes energy-dependent. The delay between the arrival times of two photons emitted simultaneously by a source at redshift $z$ with energies $E_h > E_l$ then reads:
\begin{equation}
\label{eq:timez5}
\Delta t_n \simeq \pm\,\frac{n+1}{2}\,\frac{E_h^n - E_l^n}{H_0 E_{QG,n}^n}\ \kappa_n(z),
\end{equation}
where $\kappa_n(z)$ is a parameter encoding the dependence to the distance of the source. In the case of pulsars located within our Galaxy, this function is the euclidian distance. Two expressions for $\kappa_n(z)$ are compared in this work obtained from a pure Lorentz invariance violation framework (noted hereafter J\&P) \citep{Jacob2008} and from the Doubly Special Relativity (DSR) approach \citep{Rosati2015}.

The formalism of Equation~(\ref{eq:timez5}) neglects contributions from source intrinsic effects generating time delays from emission mechanisms (see e.g. \citep{Perennes2020}). As intrinsic delays are not expected to depend on the distance, it should be possible to seperate between intrinsic and propagation effects by combining several sources at different distances. From Equation~(\ref{eq:timez5}), another parameter $\lambda_n$ can be defined as
\begin{equation}
\label{eq:lambda}
\lambda_n \equiv \frac{\Delta t_n}{\Delta E_n\ \kappa_n(z)} = \pm \frac{n+1}{2 H_0\ E^{n}_{QG,n}},
\end{equation}
using the simplified notation $\Delta E_n \equiv E_h^n - E_l^n$. This parameter has the advantage to be distance-independent and is therefore suitable for a multi-source analysis.

It is essential to perform LIV studies on a large population of objects. We present here analysis tools dedicated to population studies for the search of LIV-induced time delays with the aim of producing robust constraints on QG effects. These tools have been designed to combine for the first time the data obtained with the three major Imaging Atmospheric Cerenkov Telescope (IACT) experiments, H.E.S.S.\,\footnote{\textit{High Energy Stereoscopic System}, \url{https://www.mpi-hd.mpg.de/hfm/HESS/}}, MAGIC\,\footnote{\textit{Major Atmospheric Gamma Imaging Cherenkov}, \url{https://magic.mpp.mpg.de}} and VERITAS\,\footnote{\textit{Very Energetic Radiation Imaging Telescope Array System}, \url{https://veritas.sao.arizona.edu}}, while taking into account the distance dependence of the LIV-induced time-lag.

We will first present the two lag-distance models used in this analysis, followed by a description of the method used to compute and combine the likelihoods to measure time-lag parameter $\lambda_n$ as well as systematics treatment. The method is then tested and calibrated on simulated datasets based on several representative sources observed at TeV energies, followed by an evaluation of statistical and systematics errors. Finally, the results as well as the impact of distance dependence and systematics will be given and discussed.

\section{Distance dependence on time delays}
\label{sec:z-dep}
Amongst the two distance-lag models we consider, the one proposed by Jacob and Piran \citep[]{Jacob2008} where Lorentz invariance is explicitly broken in a specific way has already been extensively used in experimental analyses constraining in-vacuo dispersion. In this approach, parameter $\kappa_n(z)$ is expressed as:
\begin{equation}
\label{eq:kappaliv}
\kappa^{\mathrm{J\&P}}_n(z) \equiv  \int_0^z \frac{(1+z')^n}{\sqrt{\Omega_m\,(1+z')^3 + \Omega_\Lambda}}\ dz',
\end{equation}
where $H(z) = H_0 \sqrt{\Omega_m\,(1+z)^3 + \Omega_\Lambda}$ is the Hubble parameter, $H_0 = 67.74\pm0.46\ \mathrm{km\,s}^{-1}\,\mathrm{Mpc}^{-1} = (2.20\pm0.02)\times10^{-18}\ \mathrm{s}^\mathrm{-1}$, $\Omega_m = 0.3089\pm0.0062$ and $\Omega_\Lambda = 0.6911\pm0.0062$ \citep{Ade2016}.

The second model follows the Deformed Special Relativity (DSR) approach where Poincar\'e symmetries are modified in order to preserve the invariance of Equation~(\ref{eq:disprel1}) under relativistic transformations and leads to a new expression for $\kappa_n(z)$ \citep{Rosati2015}: 
\begin{equation}
\label{eq:kappadsr}
\kappa^{\mathrm{DSR}}_n(z) \equiv  \int_0^z \frac{h ^{2n}(z') dz'}{(1+z')^n\, \sqrt{\Omega_m\,(1+z')^3 + \Omega_\Lambda}},
\end{equation}
with
\begin{equation}
h (z') \equiv  1+ z' - \sqrt{\Omega_m\,(1+z')^3 + \Omega_\Lambda} \int_0^{z'} \frac{dz''}{\sqrt{\Omega_m\,(1+z'')^3 + \Omega_\Lambda}}.
\end{equation}

\begin{figure}[h!]
\vspace{-0.25in}
    \centering
    \includegraphics[width=3in]{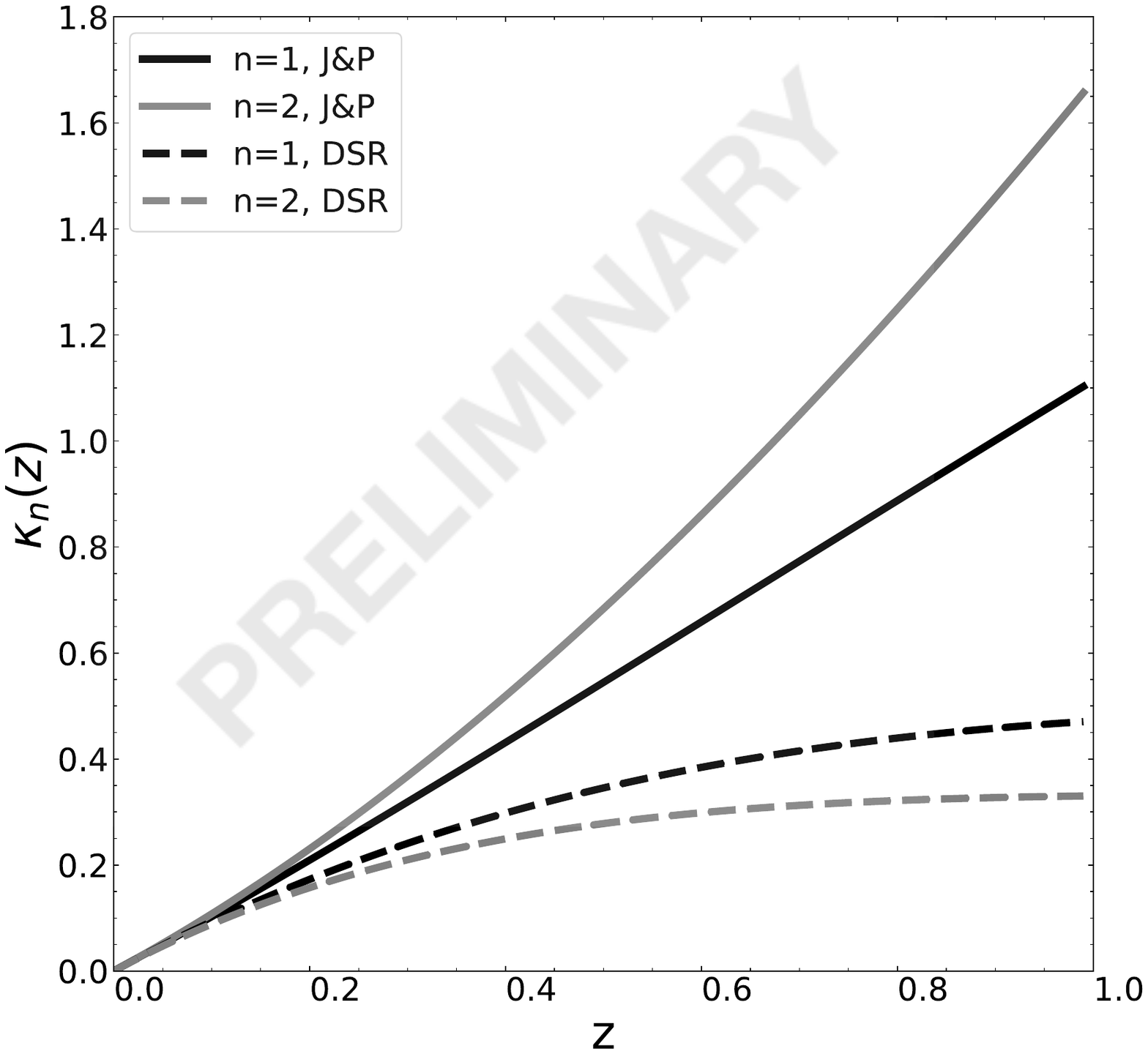}
      \vspace{-0.15in}  
    \caption{Parameter $\kappa$ for $n=1$ (black) and $n=2$ (gray) in the J\&P case (solid line) and in the DSR case (dashed line).}
    \label{fig:kappaz}
\end{figure}

Figure~\ref{fig:kappaz} shows functions $\kappa^{\mathrm{DSR}}$ and $\kappa^{\mathrm{J\&P}}$ as a function of redshift for $n=1$ and $n=2$. These two approaches appear to agree at small redshift, where $\kappa_n(z) \rightarrow d\,H_0/c$ with $d$ the euclidean distance to the source. Therefore local sources such as pulsars will give the same constraints on $E_{QG,n}$ for both J\&P and DSR cases. However $\kappa^{\mathrm{J\&P}}_n(z)$ and $\kappa^{\mathrm{DSR}}_n(z)$ significanlty diverge at large redshift leading to consistently different limits on $E_{QG,n}$. This notation $E_{QG,n}$ reflects the fact that LIV analyses have different sensitivities for linear and quadratic effects.  

\section{Methodology}
\label{sec:methodo}

\subsection{The maxium likelihood method}
\label{sec:ML}
The maximum likelihood (ML) method has been chosen to search for time delays and extract limits on $E_{QG,n}$ as it provides a straightforward way to combine analyses of multiple sources. This method requires to define a probability density function (PDF) describing the probability to observe photons at a given arrival time with a given energy. 

Parametrisations of the source true energy spectrum $\Gamma(E_t)$ and the emitted photon distribution $C(t)$ are derived from a sub-dataset of low energy photons (i.e. not significantly affected by LIV).
We define the following function for signal events
\begin{equation}
F_{s}(E_{t},t;\lambda_n) = A(E_{t}, \vec{\varepsilon}) M(E_t,E_m) \times \Gamma_{s}(E_t)C_{s}\left(t-\lambda_n \kappa_n E_t^n\right),
\end{equation}
with $A(E_{t},\vec{\varepsilon})$ the effective area and $M(E_t,E_m)$ the energy resolution, $\vec{\varepsilon}$ a set of factors encoding observation conditions and event reconstruction, and $E_m$ the measured energy. A similar function for background events which are not affected by LIV propagation effects reads:
\begin{equation}
F_{b,k}(E_{t},t) = A(E_{t}, \vec{\varepsilon}) M(E_t,E_m) \times \Gamma_{b,k}(E_t)C_{b,k}(t)
\end{equation}
with $k$ the background types (hadrons or baseline photons). The full PDF is then written as:
\begin{equation}
\frac{dP}{dE_m dt}= \frac{\int F_{s}(E_{t},t;\lambda_n) dE_{t}}{\iiint F_{s}(E_{t},t;\lambda_n) dE_t dE_m dt} + \sum_k \frac{\int F_{b,k}(E_{t},t) dE_{t}}{\iiint F_{b,k}(E_{t},t) dE_t dE_m dt}.
\end{equation}

The log-likelihood for each source $L_{S}$ is obtained by summing the log-likelihood of all the events:
\begin{equation}\label{eq:LikelihoodData}
L_{S}(\lambda_n) = \sum_{\mathrm{all\ events}} \log\left(\frac{dP}{dE_m dt}(E_{m,i},t_{i});\lambda_n\right).
\end{equation}
Maximising $L_{S}$ provides the best estimate of $\lambda_n$, along with confidence levels and lower limits on $E_{QG,n}$. The combination of multiple sources is then simply given by the sum of their individual log-likelihood:
\begin{equation}\label{eq:combination of likelihood}
L_{comb}(\lambda_n) = \sum_{\mathrm{all\ sources}} L_{S}(\lambda_n).
\end{equation}

\subsection{Statistical and systematic uncertainties}
\label{sec:syst}
Both statistical and systematic uncertainties are propagated in the final result with profile likelihoods which reads
\begin{multline}
L(\lambda_n,\vec{\theta}) = L_\mathrm{data}(\lambda_n,\vec{\theta}) + L_\mathrm{template}(\vec{\theta}_\mathrm{C}) + L_\mathrm{\gamma}(\theta_\mathrm{\gamma}) + 
L_\mathrm{BP}(\vec{\theta}_\mathrm{BP}) + L_\mathrm{ES}(\theta_\mathrm{ES}) + L_\mathrm{z}(\theta_\mathrm{z}).
\end{multline}
$\vec{\theta}$ is the vector of all nuisance parameters including the parameters of the light curve analytic parameterization $\vec{\theta}_\mathrm{C}$, the power law index of signal events spectrum $\theta_\mathrm{\gamma}$, the ratio of signal and of background event numbers to the total number of events provided by observatories $\vec{\theta}_\mathrm{BP}$, the energy scale also provided by observatories $\theta_\mathrm{ES}$, and the distance $\theta_\mathrm{z}$.

Except for $L_\mathrm{template}$, a normal distribution is assumed allowing for profile likelihoods to be defined as simple $\chi^2$ functions:
\begin{equation}
L_{\mathrm{x}}(\vec{\theta}_{\mathrm{x}}) = \sum_{i} \frac{(\theta_{\mathrm{x},i} - \bar{\theta}_{\mathrm{x},i})^2}{2\sigma^2_{\theta_{\mathrm{x},i}}},
\end{equation}
where $\sigma^2_\theta$ is the uncertainty of the nuisance parameter $\theta$, and $x$ denotes the various types of systematics.

\section{Simulations and calibration}
\label{sec:simu}

\subsection{Simulations}
The sources selected for this work are listed in Table~\ref{tab:simulation} together with their parametrisation. They have been chosen to form a representative sample which includes 3~AGNs, 2~PSRs and 1~GRB where in particular signal to background ratios, light curve shapes and distance significantly differ from one source to another. Furthermore, each source and each observation has its own set of IRF which were kindly provided by H.E.S.S., MAGIC and VERITAS collaborations. Simulated data sets are produced via Monte Carlo (MC) simulations following the specified parametrisations. 

A delay $\lambda_n^{\mathrm{inj}}$ is injected in the simulations and minimising the log-likelihood provides the most probable delay $\lambda_n^{\mathrm{rec}}$ reconstructed by the method, together with lower and upper bounds for a given confidence interval. By generating multiple MC following the same parametrisation with the same $\lambda_n^{\mathrm{inj}}$, we obtain a distribution of the reconstructed $\lambda_n^{\mathrm{rec}}$, lower and upper bounds. This distribution is Gaussian when the source light curve is symmetric, while an asymmetric light curve leads to a asymmetric Gaussian distribution. In the case of abnormally low statistics, the distribution should follow a Poisson law. 

\begin{table*}[h!]
\vspace{-0.2in} 
      \caption{Simulation settings for the individual sources.}
       \label{tab:simulation}
      \vspace{-0.25in} 
\begin{center}
\scriptsize
        \begin{tabular}{llllllll} 
        \hline
        \hline
        Source & Energy Range & Spectral index & Lightcurve shape & Number of events$^{\dagger}$  & Background proportion & Ref. \\
        & (TeV) & $\Gamma_s$, $\Gamma_{b,k}$ & & likelihood, template & hadronic, baseline & \\
        
        \hline
        GRB 190114C & $0.3$ - $2$ & $5.43$, $3.46$ & Curved power law & $726$, - & $0.055$, $0$ & \citep{Acciari2020} \\
        PG 1553+113 & $0.4$ - $0.8$ & $4.8$, $4.8$ & Double Gauss  & $72$, $82$ & $0.29$, $0.15$ & \citep{Abramowski2015}  \\ 
        Mrk 501 & $0.25$ - $11$ & $2.2$, $2.2$ & Single Gauss & $800$, - & $0.39$, $0.$ & \citep{Martinez2009}\\ 
        PKS 2155-304 & $0.28$ - $4$ & $3.46$, $3.32$ & 5 Asymmetric Gauss & $2965$, $561$ & $0.$, $0.02$ & \citep{Abramowski2011} \\ 
        Crab (M) & $0.4$ - $7$ & $2.8$, $2.47$ & Single Gauss + Baseline & $14869$, - &  $0.$, $0.961$ & \citep{Gaug2017} \\
        Crab (V) & $0.2$ - $10$ &  $3.25$, $2.47$ & Single Gauss + Baseline & $22764$, - &  $0.$, $0.964$ & \citep{Zitzer2013}\\
        Vela & $0.06$ - $0.15$ & $3.9$, 1.75 & Asymetric Lorentzian  & $3956$, - & $0.$, $0.998$ & \citep{Chretien2015} \\ \hline
        \vspace{-0.3in} 
        \end{tabular}
      
\end{center}
$^{\dagger}$ Number of photons considered when computing the likelihood, \textit{i.e.} excluding the ones used for template determination.
\end{table*}

\subsection{Calibration}
\label{sec:calib}
\begin{figure}
\hspace{-0.25in}
\begin{minipage}[t]{0.5\textwidth}
\centering
\includegraphics[width=3.07in]{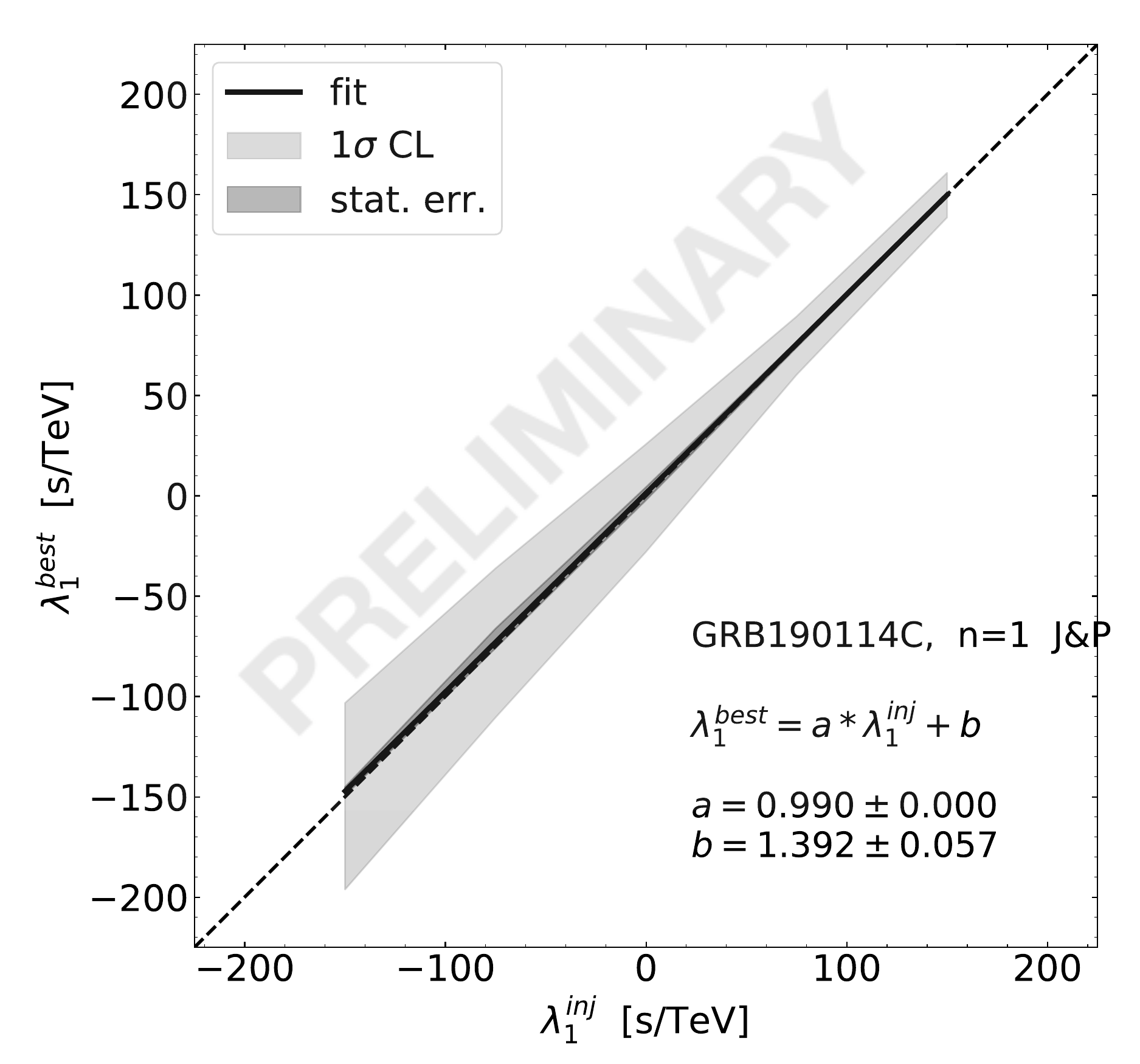}
\label{fig:calib}
\end{minipage}
\begin{minipage}[t]{0.5\textwidth}
\centering
\includegraphics[width=3in]{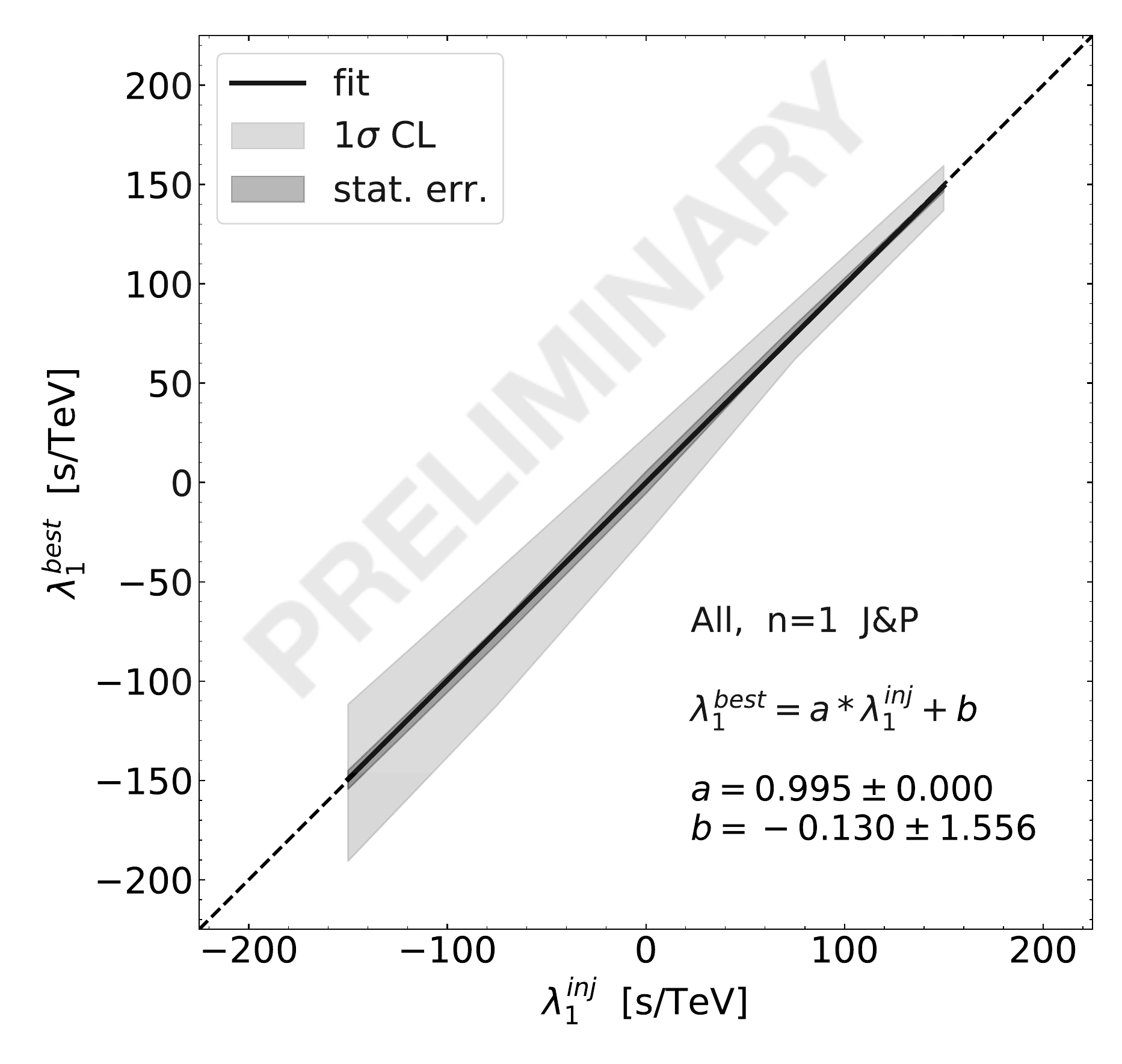}
\label{fig:calib2}
\end{minipage}
\vspace{-0.3in}
\caption{Calibration diagrams for GRB 190114C and all sources combined (linear, J\&P) showing the lag reconstructed by the method $\lambda_n^{\mathrm{rec}}$ as function of the lag injected in the simulaitons $\lambda_n^{\mathrm{inj}}$. The data points are fitted with a linear regression (solid line, $y=ax+b$), with the statistical uncertainty (dark gray area) and standard deviation (light gray area) of the $\lambda_n^{\mathrm{rec}}$ distribution.}
\end{figure}

To ensure reliable reconstructed lags $\lambda_n^{\mathrm{rec}}$, the method has been carefully calibrated. Two diagrams showing $\lambda_n^{\mathrm{rec}}$ against the injected lag $\lambda_n^{\mathrm{inj}}$ can be seen in Figure~\hyperref[fig:calib]{2} for GRB190114C and all sources combined ($n=1$, J\&P). The data points are fitted with a linear function (black line).

Although only two examples are shown here, they have been produced for all individual sources and a set of combination, for both linear and quadratic cases, as well as both J\&P and DSR models. All calibration curves fall below 8\% deviation from this ideal line, the worst case scenario is obtained with the Vela pulsar which is the closest source in the group and for which signal to background ratio is low.

\section{Results, discussion and prospects}
\label{sec:res}

\begin{figure}[h!]
\begin{minipage}[t]{\dimexpr.5\textwidth-1em}
\vspace{-0.05in}
\includegraphics[width=1\textwidth]{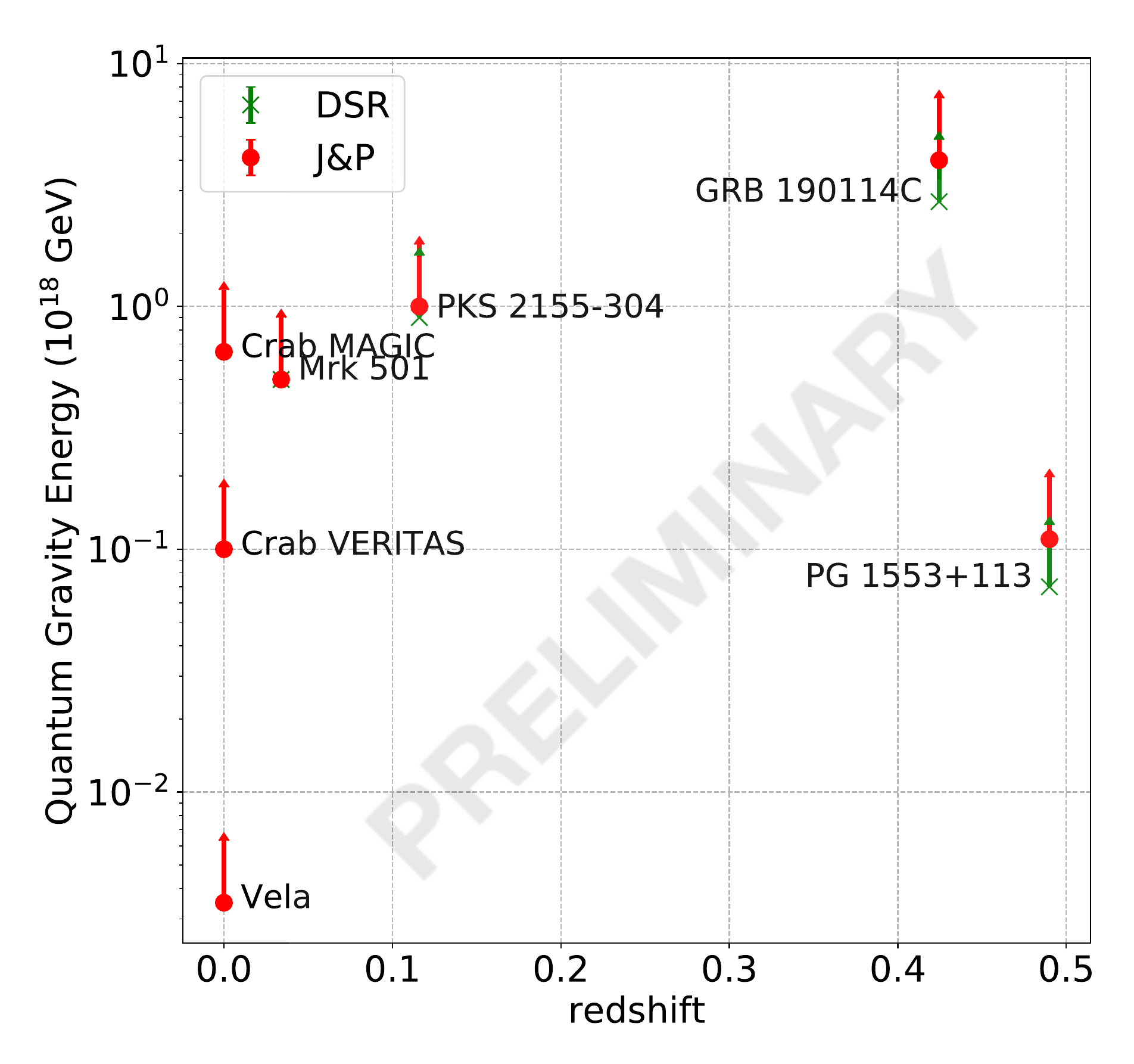}
\vspace{-0.1in}
\caption{$E_{QG,n}$ upper limits obtained in this work accounting for systematics for J\&P and DSR linear cases.}
\label{fig:limZ}
\end{minipage}
\hfill
\begin{minipage}[t]{\dimexpr.5\textwidth-1em}
\vspace{-0.05in}
\includegraphics[width=1\textwidth]{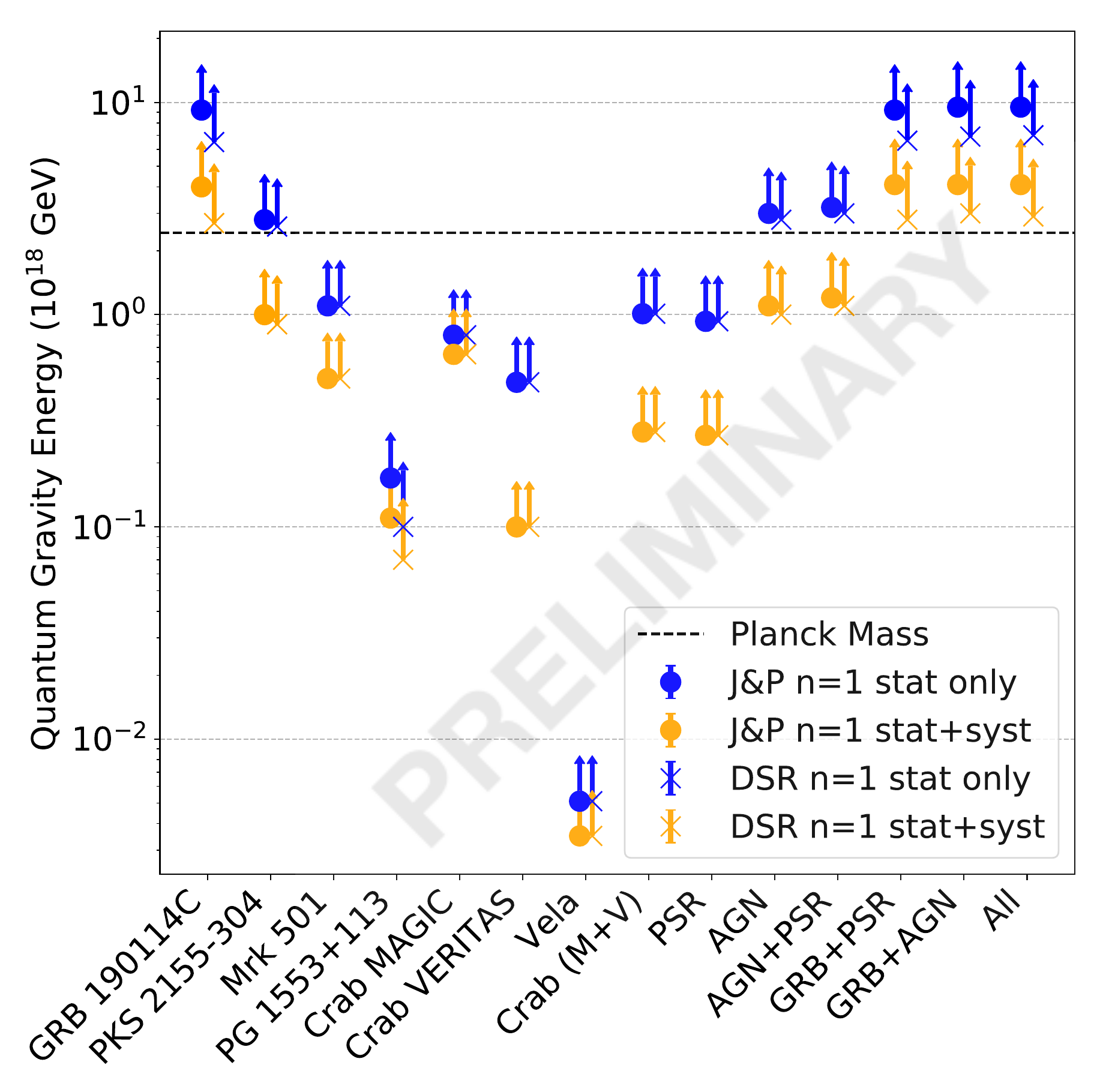}
\vspace{-0.35in}
\caption{$E_{QG,n}$ upper limits obtained in this work with and without accounting for systematics for both J\&P and DSR linear cases.}
\label{fig:lim}
\end{minipage}
\end{figure}

A summary of $E_{QG,n}$ limits for the linear case can be seen in Figure~\ref{fig:lim}. The GRB~190114C appears as the most constraining source due to its characteristics (distance, energy range, statistics, variability) especially favorable for LIV studies, and dominates the combinations when included in the sample. AGNs are the next most contraining sources. PKS~2155-304 dominates in the linear case thanks to its redshift and statistics while Mkn~501 dominates the quadratic case thanks to its wide energy range. PG~1553+113 cannot compete due to its very low statistics and small energy range, despite its higher redshift. Finally, pulsars are the least constraining sources, primarily because of their very small distance. Consequently, Vela is the least constraining source and leads to the poorest limits on $E_{QG,n}$, barely contributing to combinations. They are however the only sources independent of lag-distance models.

Regarding the impact of DSR and J\&P models on $E_{QG,n}$ limits, a summary for the linear case can be found in Figure~\ref{fig:limZ}. The differences start to become tangible for high redshift sources such as GRB~190114C or PG~1553+311 as could have been expected. On the one hand, the J\&P model appears to emphasize the impact of large distance sources on $E_{QG,n}$ limits, further establishing the GRB's dominance over the other types of sources. On the other hand the DSR model tends to balance sources' contribution such that their combination leads to a significant improvement on $E_{QG,n}$ limits.

Individual sources appear to be dominated by systematics from the light curve template in the linear case, and the precision on the energy distribution in the quadratic case, except for the GRB 190114C which is dominated by the power law index. A summary for the linear case can be seen in Figure~\ref{fig:lim}. Combinations are dominated by the most stringent source in the sample and its dominant systematic. Results have been reported with and without accounting for systematic uncertainties to show how great an impact they can have on $E_{QG,n}$ upper limits, often dividing them by a factor~$\gtrsim 2$. Overall, the simulated datasets are in good agreement with the actual data. The observed differences most likely arise from the combined differences in IRF and systematics treatment, but also from the thousand Monte Carlo simulations used for this study as opposed to the one measured lightcurve used in previous papers.

The next steps will be to analyse new datasets and combine larger samples to further constrain quantum energy scales, and introduce it to analysis pipelines used for the future Cherenkov Telescope Array (CTA) observatory, the next generation of IACTs.

\begin{table}[h!]
\begin{center}
\caption{95\% CL limits obtained for individual objects and combinations.}
\label{tab:combined_res}
\scriptsize
\begin{tabular}{lrlrlrlrlrlrl}
\hline
\hline
\multirow{3}{*}{Source} & \multicolumn{4}{c}{$E_{\mathrm{QG},1}$ ($10^{18}$ GeV)} & \multicolumn{4}{c}{$E_{\mathrm{QG},2}$ ($10^{10}$ GeV)} \\
    & \multicolumn{2}{c}{J\&P}  & \multicolumn{2}{c}{DSR} & \multicolumn{2}{c}{J\&P} & \multicolumn{2}{c}{DSR} \\
    & w/o syst. & w/ syst. & w/o syst. & w/ syst. & w/o syst. & w/ syst. & w/o syst. & w/ syst.  \\
\hline
GRB 190114C & 9.2 & 4.0 & 6.5 & 2.7 & 14.2 & 8.3 & 9.5 & 5.8 \\
PKS  2155-304 & 2.8 & 1.0 & 2.6 & 0.9 & 8.2 & 6.2 & 7.2 & 5.5 \\
Mrk~501 & 1.1 & 0.5 & 1.1 & 0.5 & 9.6 & 7.1 & 9.3 & 6.9 \\
PG 1553+113 & 0.17 & 0.11 & 0.10 & 0.07 & 1.3 & 1.0 & 0.87 & 0.68 \\
Crab (M)   & 0.80 & 0.65 & - & - & 3.0 & 2.5 & - & - \\
Crab (V)   & 0.48 & 0.10 & - & - & 1.5 & 0.94 & - & - \\
Vela  & $5.1 \times 10^{-3}$ & $3.5 \times 10^{-3}$ & - & - & $5.6 \times 10^{-2}$ & $5.5 \times 10^{-2}$ & - & - \\
\hline
Crab (M+V)  & 1.0 & 0.28 & - & - & 3.3 & 2.6 & - & - \\
PSR & 1.0 & 0.28 & - & - & 3.3 & 2.8 & - & - \\
AGN & 3.0 & 1.1 & 2.8 & 1.0 & 10.8 & 8.3 & 10.5 & 7.9 \\
AGN+PSR & 3.2 & 1.2 & 3.0 & 1.1 & 10.6 & 8.5 & 10.1 & 8.3 \\
GRB+PSR & 9.2 & 4.1 & 6.6 & 2.8 & 14.3 & 9.2 & 9.1 & 7.0 \\
GRB+AGN & 9.5 & 4.1 & 6.9 & 3.0 & 14.5 & 9.7 & 11.4 & 8.2 \\
\hline
All combined & 9.5 & 4.1 & 7.0 & 2.9 & 14.4 & 9.7 & 11.1 & 8.4 \\
\hline
\vspace{-0.3in}
\end{tabular}
\end{center}
\end{table}

%
%
%


\begin{thebibliography}{99}
\bibitem[Abramowski et al.(2011)]{Abramowski2011} Abramowski, A. et al. (H.E.S.S. Collaboration), 2011, Astropart. Phys, 34, 738
\bibitem[Abramowski et al.(2015)]{Abramowski2015} Abramowski, A. et al. (H.E.S.S. Collaboration), 2015, ApJ, 802, 65
\bibitem[Acciari et al.(2020)]{Acciari2020} Acciari, V.~A., Ansoldi, S., Antonelli, L.~A., et al.\ 2020, Phys. Rev. Lett. 125, 021301.
\bibitem[Ade et al.(2016)]{Ade2016} Ade, P. A. R. et al. (Planck Collaboration ), 2016, A\&A, 594, A13
\bibitem[Amelino-Camelia et al.(1998)]{Amelino1998} Amelino-Camelia, G. et al., 1998, Nature, 393, 763
\bibitem[Brun \& Rademakers(1997)]{ROOT} Brun, R. \& Rademakers, F., 1997, Nucl. Inst. \& Meth. in Phys. Res. A, 389, 81
\bibitem[Chretien et al.(2015)]{Chretien2015} Chretien, M. et al., 2015, Proceedings of ICRC 2015
\bibitem[Gambini \& Pullin(1999)]{Gamibini1999} Gambini, R. \& Pullin, J., 1999, PhRvD, 59, 124021
\bibitem[Gaug et al.(2017)]{Gaug2017} Gaug, M., Garrido, D., \& MAGIC Collaboration\ 2017, Proceedings of ICRC 2017
\bibitem[Jacob \& Piran(2008)]{Jacob2008} Jacob, U., \& Piran, T., 2008, JCAP, 01, 031
\bibitem[Jacobson et al.(2006)]{Liberati2006} Jacobson, T. et al., 2006, Annals Phys.,  321, 150 
\bibitem[Martinez \& Errando(2009)]{Martinez2009} Mart{\'{\i}}nez, M. \& Errando, M., 2009, Astropart. Phys, 31, 226
\bibitem[Mavromatos(2010)]{Mavromatos2010} Mavromatos, N. E., 2010, IJMPA, 25, 5409
\bibitem[Perennes et al.(2020)]{Perennes2020} Perennes, C. , Sol, H. and Bolmont, J., 2020, A\&A, 633, A143
\bibitem[Rosati et al.(2015)]{Rosati2015} Rosati, G. et al., 2015, PhRvD, 92,  124042 
\bibitem[Zitzer et al.(2013)]{Zitzer2013} Zitzer, B. et al., 2013, Proceedings of ICRC 2013


\end{thebibliography}
\end{document}